# Electrophilicity Equalization Principle

*Pratim Kumar Chattaraj\*, Santanab Giri and Soma Duley*

Department of Chemistry and Center for Theoretical Studies,

Indian Institute of Technology, Kharagpur, 721302, India

E-mail- pkc@chem.iitkgp.ernet.in

**ABSTRACT**   A new electronic structure principle, viz., the principle of electrophilicity equalization is proposed. An analytical justification as well as a numerical support for the same is provided.

**SECTION**     Molecular Structure, Quantum Chemistry, General Theory

In this Letter, we propose a new electronic structure principle, viz., the principle of electrophilicity equalization and provide an analytical justification as well as a numerical support thereof. It states that, "During molecule formation, like electronegativity and hardness, electrophilicity gets equalized".

Conceptual density functional theory[1-3] provides quantitative definitions for popular qualitative chemical concepts and theoretical bases for associated electronic structure principles. Chemical potential ($\mu$) (the Lagrange multiplier associated with the normalization constraint and the negative of the electronegativity[4] ($\chi$)) and hardness[5] ($\eta$) are given by the following first - and the second – order derivatives respectively,

$$\mu = -\chi = \left(\frac{\partial E}{\partial N}\right)_{v(\mathbf{r})} \qquad (1)$$

and

$$\eta = \left(\frac{\partial^2 E}{\partial N^2}\right)_{v(\mathbf{r})} \quad , \tag{2}$$

for an N – electron system with energy E and external potential υ(**r**). Using a finite difference approximation one may approximate them in terms of the ionization potential (I) and electron affinity (A) as[1,6]

$$\chi = -\mu \approx \left(\frac{I+A}{2}\right) \tag{3}$$

and

$$\eta = I - A \quad . \tag{4}$$

Electrophilicity (ω) of the same system may be defined as[7-9]

$$\omega = \left(\frac{\mu^2}{2\eta}\right) = \left(\frac{\chi^2}{2\eta}\right) \quad . \tag{5}$$

Sanderson's electronegativity equalization principle states that[10,11], "When two or more atoms initially different in electronegativity combine chemically their electronegativities become equalized in the molecule. The equalization of electronegativity occurs through the adjustment of the polarities of the bonds which is pictured as resulting in a partial charge on each atom. That is, electron loss causes increase, and electron gain decrease in electronegativity". Final molecular electronegativity $(\chi_{GM})$ may be expressed approximately as the geometric mean of the electronegativities of the isolated atoms[12], i.e.,

$$\chi_{GM} \approx \left(\prod_{k=1}^{P} \chi_k\right)^{\frac{1}{P}} \tag{6}$$

if the molecule contains P atoms (same and / or different) and $\{\chi_k, k = 1, 2, ......., P\}$ denote their isolated atom electronegativities.

It has been observed[13] that the ratio of hardness and electronegativity is roughly a constant for atoms (at least for the atoms belonging to the same group in the periodic table[14]) and molecules. Based on this fact and eq (6) it may be shown[15] that

$$\eta_{GM} \approx \left( \prod_{k=1}^{P} \eta_k \right)^{\frac{1}{P}} \qquad (7)$$

where $\{\eta_k, k = 1, 2, \ldots, P\}$ refer to the corresponding isolated atom hardness values. Therefore hardness also gets equalized like electronegativity. Equality between local and global hardness values reaffirms this fact albeit making the definition of the former ambiguous[16].

In this Letter, we show that the validity of the above two equalization principles tantamounts to that of the electrophilicity equalization principle. Expressing $\omega$ of a molecule ($\omega_{GM}$) as

$$\omega_{GM} = \frac{\chi_{GM}^2}{2\eta_{GM}} \qquad (8)$$

we have

$$\omega_{GM} = \frac{\left( \prod_{k=1}^{P} \chi_k \right)^{\frac{2}{P}}}{2 \left( \prod_{k=1}^{P} \eta_k \right)^{\frac{1}{P}}}$$

$$= \frac{1}{2} \left( \prod_{k=1}^{P} \frac{\chi_k^2}{\eta_k} \right)^{\frac{1}{P}}$$

$$= \frac{1}{2} \left( 2^P \prod_{k=1}^{P} \omega_k \right)^{\frac{1}{P}}$$

$$= \left( \prod_{k=1}^{P} \omega_k \right)^{\frac{1}{P}}, \qquad (9)$$

$\left\{ \omega_k = \frac{\chi_k^2}{2\eta_k}, k = 1, 2, \ldots, P \right\}$ being the electrophilicities of the isolated atoms.

Therefore the electrophilicity gets equalized during molecule formation and the final equalized electrophilicity may be expressed as the geometric mean of the corresponding isolated atom values. When an electrophile interacts with a nucleophile the electrophilicity of the former gets reduced (via electronic charge transfer and/or other related processes, from the nucleophile to the electrophile) and that of the latter gets increased until they get equalized to a value somewhere in between the two (roughly the geometric mean).

An important outcome of this result is that the local electrophilicity[17] may be considered to be constant everywhere and is equal to its global variant.

Numerical calculations are performed at the B3LYP/6-311+G** level of theory to calculate the energies of some selected atoms and their cations/anions to obtain the $I$ and $A$ values of the atoms using a $\Delta SCF$ technique. Geometries of the corresponding diatomic molecules are optimized at the same level of theory using a Gaussian-suite of program[18] and the energies of the related cations/anions are obtained through single point calculations using the geometries of the associated neutral molecules. Table 1 presents the energy ($E$), ionization potential ($I$), electron affinity ($A$), electronegativity ($\chi$), hardness ($\eta$), electrophilicity ($\omega$) and the last three quantities approximated as the respective geometric means $\left(\lambda_{GM} = \sqrt{\lambda_A \cdot \lambda_B} \; ; \; \lambda \equiv \chi, \eta, \omega\right)$ and the ratios of the two sets of values ($\lambda_R = \dfrac{\lambda}{\lambda_{GM}} \; ; \; \lambda \equiv \chi, \eta, \omega$). Table 2 reports these quantities for some selected molecules, calculated from the experimental $I$ and $A$ values[13]. Geometric mean principle for $\omega$ is obeyed very well wherever the corresponding principles for $\chi$ and $\eta$ are obeyed properly.

In conclusion, the electrophilicity gets equalized during molecule formation, like the electronegativity and the hardness. The final electrophilicity is roughly given by the geometric mean of the corresponding isolated atom values. Like local hardness, local electrophilicity may be taken as the same as the global electrophilicity.

ACKNOWLEDGMENT

We thank CSIR, New Delhi for financial assistance.

Table 1. Energy (*E*, a.u.), ionization potential (*I*, a.u.), electron affinity (*A*, a.u.), electronegativity ($\chi$, a.u.), hardness ($\eta$, a.u.) and electrophilicity ($\omega$, a.u.) values of some selected diatomic molecules, calculated at the B3LYP/6-311+G** level of theory.

| Molecules | E | I | A | $\chi$ | $\chi_{GM}$ | $\chi_R$ | $\eta$ | $\eta_{GM}$ | $\eta_R$ | $\omega$ | $\omega_{GM}$ | $\omega_R$ |
|---|---|---|---|---|---|---|---|---|---|---|---|---|
| LiF | -107.46822 | 0.432 | 0.018 | 0.225 | 0.228 | 0.987 | 0.414 | 0.350 | 1.183 | 0.061 | 0.074 | 0.824 |
| LiCl | -467.83380 | 0.372 | 0.025 | 0.199 | 0.199 | 1.00 | 0.348 | 0.280 | 1.243 | 0.057 | 0.070 | 0.814 |
| LiBr | -2581.75601 | 0.351 | 0.027 | 0.189 | 0.190 | 0.995 | 0.324 | 0.264 | 1.227 | 0.055 | 0.069 | 0.797 |
| NaF | -262.22188 | 0.381 | 0.025 | 0.203 | 0.225 | 0.902 | 0.356 | 0.342 | 1.041 | 0.058 | 0.074 | 0.784 |
| NaCl | -622.60101 | 0.344 | 0.032 | 0.188 | 0.196 | 0.959 | 0.312 | 0.274 | 1.139 | 0.057 | 0.070 | 0.814 |
| NaBr | -2736.52691 | 0.327 | 0.033 | 0.18 | 0.188 | 0.957 | 0.294 | 0.258 | 1.140 | 0.055 | 0.068 | 0.809 |
| KF | -699.87089 | 0.366 | 0.017 | 0.191 | 0.205 | 0.932 | 0.349 | 0.310 | 1.126 | 0.052 | 0.068 | 0.765 |
| KCl | -1060.24770 | 0.325 | 0.025 | 0.175 | 0.179 | 0.978 | 0.300 | 0.248 | 1.210 | 0.051 | 0.065 | 0.785 |
| KBr | -3174.17350 | 0.309 | 0.027 | 0.168 | 0.172 | 0.977 | 0.282 | 0.234 | 1.205 | 0.050 | 0.063 | 0.794 |
| BeO | -89.93401 | 0.375 | 0.082 | 0.229 | 0.243 | 0.942 | 0.293 | 0.371 | 0.790 | 0.089 | 0.079 | 1.127 |
| MgO | -275.26018 | 0.289 | 0.072 | 0.180 | 0.223 | 0.807 | 0.217 | 0.343 | 0.633 | 0.075 | 0.073 | 1.027 |
| CaO | -752.81566 | 0.258 | 0.034 | 0.146 | 0.202 | 0.723 | 0.225 | 0.301 | 0.748 | 0.047 | 0.068 | 0.691 |
| BeS | -412.92995 | 0.341 | 0.087 | 0.214 | 0.212 | 1.009 | 0.255 | 0.302 | 0.844 | 0.090 | 0.074 | 1.216 |
| MgS | -598.29821 | 0.288 | 0.079 | 0.183 | 0.195 | 0.938 | 0.210 | 0.279 | 0.753 | 0.080 | 0.068 | 1.176 |
| CaS | -1075.83022 | 0.257 | 0.051 | 0.154 | 0.177 | 0.870 | 0.206 | 0.245 | 0.841 | 0.058 | 0.064 | 0.906 |

Table 2. Electronegativity ($\chi$, eV), hardness ($\eta$, eV) and electrophilicity ($\omega$, eV) values of some selected molecules using experimental ionization potential ($I$, eV) and electron affinity ($A$, eV) values.

| Molecules | I | A | $\chi$ | $\chi_{GM}$ | $\chi_R$ | $\eta$ | $\eta_{GM}$ | $\eta_R$ | $\omega$ | $\omega_{GM}$ | $\omega_R$ |
|---|---|---|---|---|---|---|---|---|---|---|---|
| *Diatomic molecules* | | | | | | | | | | | |
| I$_2$ | 9.400 | 2.420 | 5.910 | 6.758 | 0.875 | 6.980 | 7.393 | 0.944 | 2.502 | 3.088 | 0.810 |
| BrI | 9.790 | 2.550 | 6.170 | 7.167 | 0.861 | 7.240 | 7.916 | 0.915 | 2.629 | 3.245 | 0.810 |
| S$_2$ | 9.400 | 1.660 | 5.530 | 6.217 | 0.889 | 7.740 | 8.280 | 0.935 | 1.976 | 2.334 | 0.847 |
| Br$_2$ | 10.560 | 2.600 | 6.580 | 7.602 | 0.866 | 7.960 | 8.476 | 0.939 | 2.720 | 3.409 | 0.798 |
| Cl$_2$ | 11.480 | 2.400 | 6.940 | 8.313 | 0.835 | 9.080 | 9.395 | 0.966 | 2.652 | 3.677 | 0.721 |
| P$_2$ | 9.600 | 0.650 | 5.125 | 5.615 | 0.913 | 8.950 | 9.738 | 0.919 | 1.467 | 1.619 | 0.906 |
| SO | 10.000 | 1.130 | 5.565 | 6.846 | 0.813 | 8.870 | 10.031 | 0.884 | 1.746 | 2.336 | 0.747 |
| C$_2$ | 12.600 | 3.540 | 8.070 | 6.262 | 1.289 | 9.060 | 9.988 | 0.907 | 3.594 | 1.963 | 1.831 |
| CH | 10.640 | 1.240 | 5.940 | 6.703 | 0.886 | 9.400 | 11.325 | 0.830 | 1.877 | 1.984 | 0.946 |
| CN | 14.500 | 3.820 | 9.160 | 6.747 | 1.358 | 10.680 | 12.051 | 0.886 | 3.928 | 1.889 | 2.079 |
| O$_2$ | 12.060 | 0.440 | 6.250 | 7.538 | 0.829 | 11.620 | 12.152 | 0.956 | 1.681 | 2.338 | 0.719 |
| OH | 13.180 | 1.830 | 7.505 | 7.354 | 1.021 | 11.350 | 12.492 | 0.909 | 2.481 | 2.165 | 1.146 |
| NH | 13.100 | 0.380 | 6.740 | 7.222 | 0.933 | 12.720 | 13.664 | 0.931 | 1.786 | 1.909 | 0.936 |
| F$_2$ | 15.700 | 3.080 | 9.390 | 10.410 | 0.902 | 12.620 | 14.021 | 0.900 | 3.493 | 3.864 | 0.904 |
| *Triatomic molecules* | | | | | | | | | | | |
| CS$_2$ | 10.080 | 1.000 | 5.540 | 6.232 | 0.889 | 9.080 | 8.814 | 1.030 | 1.690 | 2.203 | 0.767 |
| COS | 11.180 | 0.460 | 5.820 | 6.645 | 0.876 | 10.720 | 10.017 | 1.070 | 1.580 | 2.204 | 0.717 |
| SO$_2$ | 12.340 | 1.050 | 6.695 | 7.069 | 0.947 | 11.290 | 10.693 | 1.056 | 1.985 | 2.337 | 0.849 |
| O$_3$ | 12.670 | 1.820 | 7.245 | 7.538 | 0.961 | 10.850 | 12.152 | 0.893 | 2.419 | 2.338 | 1.035 |
| NH$_2$ | 12.800 | 0.780 | 6.790 | 7.206 | 0.942 | 12.020 | 13.384 | 0.898 | 1.918 | 1.940 | 0.989 |
| N$_2$O | 12.890 | 1.470 | 7.180 | 7.358 | 0.976 | 11.420 | 13.696 | 0.834 | 2.257 | 1.977 | 1.142 |

### Polyatomic molecules

| | | | | | | | | | | | |
|---|---|---|---|---|---|---|---|---|---|---|---|
| PBr$_3$ | 9.850 | 1.600 | 5.725 | 7.047 | 0.812 | 8.250 | 8.775 | 0.940 | 1.986 | 2.830 | 0.702 |
| PCl$_3$ | 9.910 | 0.800 | 5.355 | 7.536 | 0.711 | 9.110 | 9.480 | 0.961 | 1.574 | 2.995 | 0.526 |
| POCl$_3$ | 11.400 | 1.400 | 6.400 | 7.536 | 0.849 | 10.00 | 9.962 | 1.004 | 2.048 | 2.851 | 0.718 |
| CH$_3$I | 9.540 | 0.200 | 4.870 | 6.899 | 0.706 | 9.340 | 10.935 | 0.854 | 1.270 | 2.176 | 0.584 |
| SO$_3$ | 11.000 | 1.700 | 6.350 | 7.184 | 0.884 | 9.300 | 11.041 | 0.842 | 2.168 | 2.337 | 0.928 |
| CF$_3$I | 10.230 | 1.400 | 5.815 | 8.625 | 0.674 | 8.830 | 11.527 | 0.766 | 1.915 | 3.227 | 0.593 |
| C$_2$H$_2$ | 11.410 | 0.430 | 5.920 | 6.703 | 0.883 | 10.980 | 11.325 | 0.970 | 1.596 | 1.984 | 0.804 |
| CF$_3$Br | 11.820 | 0.910 | 6.365 | 8.831 | 0.721 | 10.910 | 11.847 | 0.921 | 1.857 | 3.291 | 0.564 |
| CH$_3$ | 9.840 | 0.080 | 4.960 | 6.935 | 0.715 | 9.760 | 12.059 | 0.809 | 1.260 | 1.994 | 0.632 |
| HNO$_3$ | 11.030 | 0.570 | 5.800 | 7.410 | 0.783 | 10.460 | 12.736 | 0.821 | 1.608 | 2.156 | 0.746 |
| SF$_6$ | 15.350 | 0.750 | 8.050 | 9.671 | 0.832 | 14.600 | 13.005 | 1.123 | 2.219 | 3.596 | 0.617 |
| C$_6$H$_5$NO$_2$ | 9.860 | 0.700 | 5.280 | 6.823 | 0.774 | 9.160 | 11.542 | 0.794 | 1.522 | 2.017 | 0.755 |
| C$_6$H$_4$O$_2$ | 9.670 | 1.890 | 5.780 | 6.758 | 0.855 | 7.780 | 11.221 | 0.693 | 2.147 | 2.035 | 1.055 |